\documentclass[12pt]{amsart}
\calclayout
\usepackage[utf8]{inputenc}
\usepackage{amsmath,amsthm,amssymb}
\usepackage{mathrsfs}
\usepackage{colonequals}

\usepackage{graphicx}
\usepackage[colorlinks=true,allcolors=blue,pdftitle={Fingerprint-like pattern formation via reaction-difussion}]{hyperref}
\usepackage[initials,nobysame,numeric,msc-links,backrefs]{amsrefs}
\usepackage{color}

\usepackage[font=small]{caption}

\title[PDE model for fingerprint-like pattern emergence]{Novel reaction-diffusion PDE model for fingerprint-like pattern emergence via the Schnakenberg mechanism}

\author[F. Sep\'ulveda-Soto]{Fabi\'an Sep\'ulveda-Soto}
\address[F. Sep\'ulveda-Soto]{Universidad de Chile, Departamento de Ingeniería Matem\'atica, Santiago, 8370459, Chile}
\email{fsepulveda@dim.uchile.cl}

\author[L. Soto-Barrios]{Lucia Soto-Barrios}
\address[L. Soto-Barrios]{Escuela de Investigaciones Policiales, PDI, Santiago, 9160000, Chile}
\email{lsotob@investigaciones.cl}

\author[C. Rom\'{a}n]{Carlos Rom\'{a}n}
\address[C. Rom\'{a}n]{Facultad de Matem\'aticas e Instituto de Ingenier\'ia Matem\'atica y Computacional, Pontificia Universidad Cat\'olica de Chile, Vicu\~na Mackenna 4860, 7820436 Macul, Santiago, Chile}
\email{carlos.roman@uc.cl}

\author[A. Osses]{Axel Osses}
\address[A. Osses]{Departamento de Ingenier\'ia Matem\'atica y Centro de Modelamiento Matem\'atico (UMI 2807 CNRS), FCFM Universidad de Chile, Casilla 170/3 - Correo 3, Santiago, Chile}
\email{axosses@dim.uchile.cl}

\date{November 4, 2025}

\begin{document}
\begin{abstract}
Fingerprint analysis and fingerprint identification have been the most widely used tools for human identification. To this day, various models have been proposed to explain how fingerprints are formed, ranging from the fibroblast model, which focuses on cell-collagen interactions, to the buckling of thin layers model, both yielding significant results. In this work, we present a reaction‑diffusion model of Schnakenberg type, featuring an anisotropic diffusion matrix that follows the ridge orientations supplied by other traditional fingerprint‑generation models, and notably yet allows minutiae---i.e. characteristic microstructures embedded in fingerprints---to emerge. The statistical analysis of the minutiae distribution in a randomly generated fingerprint collection is consistent with observations in real fingerprints.
The model can numerically generate fingerprint-like patterns corresponding to the four basic classifications---arches, ulnar loops, radial loops, and whorls---as well as a variety of derived forms. The generated patterns emerge on a convex domain that mimics the geometry of a fingertip, exhibiting the diverse types of minutiae typically analyzed in fingerprint identification and showing strong agreement with those observed in human fingerprints. This model also provides insight into how levels of certainty in human identification can be achieved when based on minutiae positions. All the algorithms are implemented in an open source software named GenCHSin.
\end{abstract}

\maketitle
\section{Introduction}\label{sec:Intro}

Fingerprints exhibit a variety of structural patterns that assist forensic experts in the task of human identification. They have become a central tool in cases where unidentified individuals need to confirm their identities. Even in modern technologies-such as obtaining a passport or personal ID, accessing bank accounts, entering restricted areas in buildings, or unlocking cellphones-fingerprints serve as the master key to everyday security. This widespread use is no coincidence; it is the result of over 200 years of observations, detailed classification, and structured analysis across diverse populations. 

Since the late 1700s, fingerprints have been studied, beginning with the German anatomist Johann C. A. Mayer, who was the first European to recognize in 1788 that fingerprints are unique to each individual \cite{mayer1788}. This recognition marked a milestone upon which the discipline of fingerprinting would later be built. In the late 1800s, Victor Balthazard explored how personal identification could aid criminal investigations. He studied the traces of hair and their characteristics, publishing with M. Lambert the first comprehensive study on the subject \cites{baltazard1,baltazard2}. Balthazard later analyzed the combinatorial properties of fingerprint characteristics and developed a statistical model, over the hypothesis that accidents on the fingerprint's ridge happen independently, that quantified the certainty of identification. His model was subsequently used by Edmond Locard to establish rules for personal identification in his manual \cite{locard1923manuel}. 

All of this foundational work was focused on the examination and comparison of microstructures embedded in fingerprints, known as minutiae. In 1892, Sir Francis Galton published his studies \cite{galton} on how the unique characteristics of fingerprints could be used for personal identification, detailing the types of structures found within these patterns. He complemented his research, particularly in the context of criminal records and the identification of unknown individuals, with the anthropometric studies of Alphonse Bertillon \cite{bertillon1893identification}. Bertillon’s method involved measuring physical features of fully developed individuals that should not change over time. However, the approach failed when two people with very similar features were found to share identical measurements, revealing the need for a truly invariable trait of the human body: fingerprints.

Galton’s studies, along with the failure of Bertillon’s anthropometric identification method, led to the recognition of three fundamental characteristics that make fingerprints suitable for standardized human identification:
\begin{enumerate} 
\item \textbf{Immutability}: Once formed, ridge's pattern organization do not change due to external factors. 
\item \textbf{Permanence}: The patterns persist throughout a person’s lifetime. 
\item \textbf{Variability}: Fingerprints exhibit unique patterns across individuals, ensuring no two are alike.
\end{enumerate}

\begin{figure}[ht]
    \centering
    \includegraphics[width=0.6\textwidth]{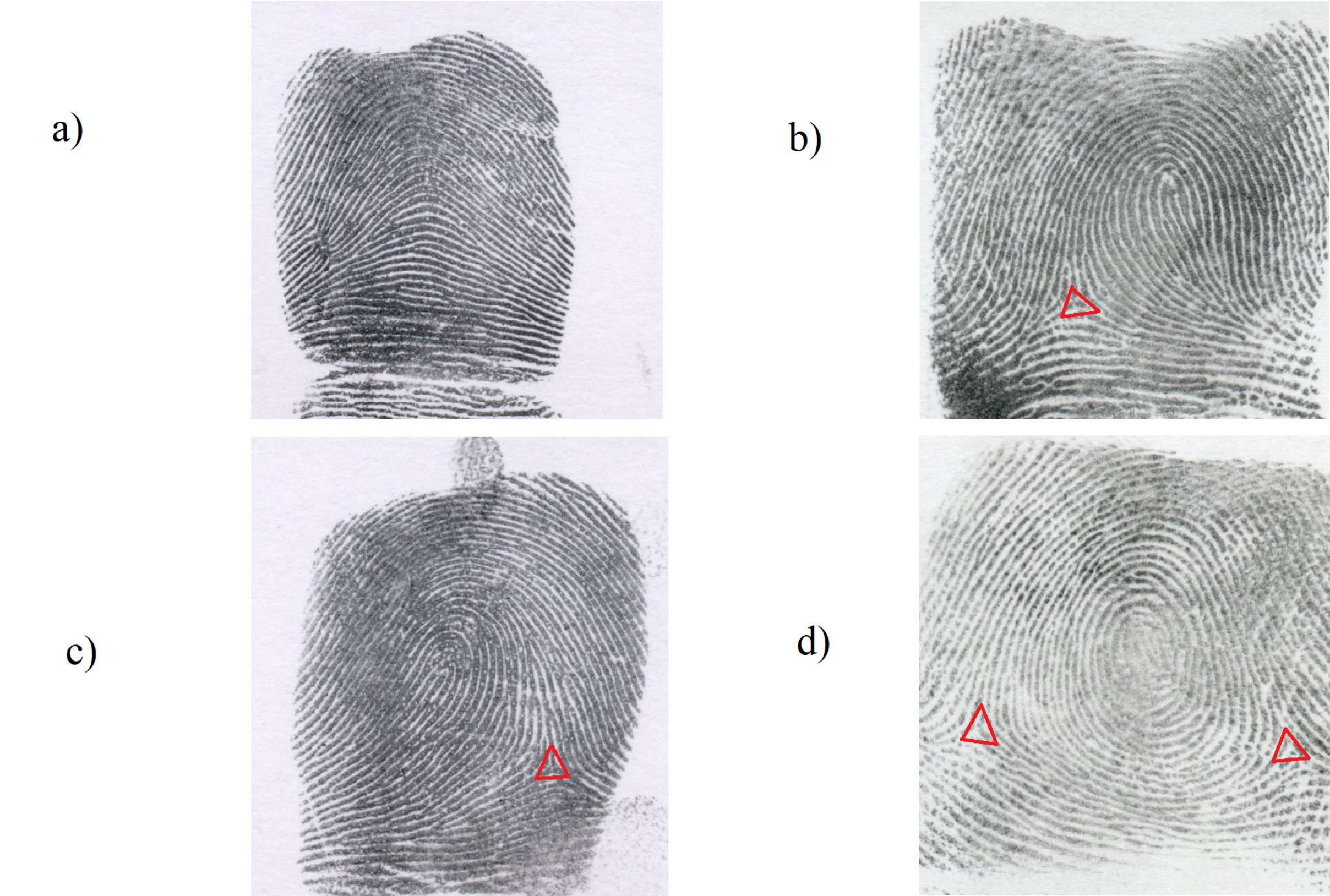}
    \caption{The fundamental classification of fingerprint patterns, introduced by Juan Vucetich, which is based on the presence or absence of a delta structure, illustrated as a red triangle in the figures. Depending on the number of deltas (or their absence), fingerprint patterns can be categorized as follows: a) Simple arch b) Simple external loop c) Simple internal loop d) Mononuclear whorl.}
    \label{Deltas}
\end{figure}

These three properties have served as a foundational basis for forensic science and law enforcement worldwide, guiding the use of fingerprint structures and their unique features for personal identification over the years. Galton’s findings and advances in fingerprint analysis were crucial to inspiring Juan Vucetich to become his apprentice. Building on these insights, Vucetich developed a process and a set of instructions for criminal identification \cite{vucetich1896instrucciones}, along with a simple classification system for the various fingerprint patterns that may appear at crime scenes \cite{vucetich1904dactiloscopia}. This basic classification, still in use today, relies on the presence or absence of a structure resembling the Greek letter $\Delta$, known as the ``delta''. This system remains the cornerstone of fingerprint classification and is illustrated in Fig.~\ref{Deltas} with several examples.

\begin{figure}[ht]
    \centering
    \includegraphics[width=0.8\textwidth]{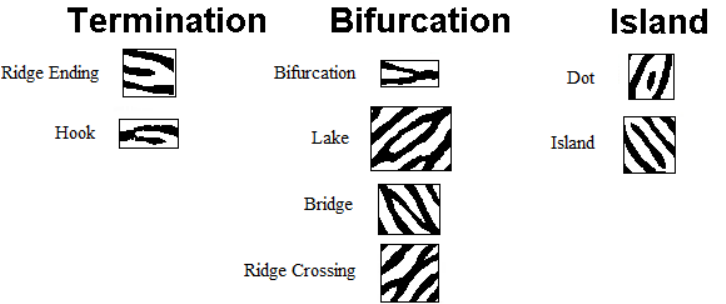}
    \caption{Classification of different minutiae based on the basic structures observed in fingerprint patterns.}
    \label{minutas}
\end{figure}

The classification assigned to a fingerprint is not sufficient for personal identification on its own. Within these patterns, multiple structures already mentioned and known as minutiae appear, which can be categorized based on different combinations of three basic features: terminations, bifurcations, and islands. These three fundamental structures can be combined to form more complex minutiae, as shown in Fig.~\ref{minutas}.

J. Vucetich’s work has become the cornerstone of modern personal identification. To determine the correspondence of a fingerprint, the direction of the ridges is first analyzed, dividing the pattern into three zones: nuclear, basal, and marginal (i.e., center, bottom, and top regions of the domain). This structural division follows the principles of his foundational classification system.
Next, a minimum number of minutiae must be considered, at least six, although the exact requirement varies by country. These minutiae, along with their relative positions within the pattern, must match those of a fingerprint stored in a database. The minimum number of required minutiae was estimated using a combinatorial rule originally devised by V. Balthazard \cites{baltazard1,baltazard2} and subsequently cited and endorsed by E. Locard \cite{locard1923manuel}, based on the size of the global population.
If the classification, ridge direction, and relative positions of the identified minutiae align with a fingerprint in the database, the fingerprint can be considered identified. This identification process has been extensively studied in various contexts \cites{Silva, kinsgston, teke2010medicina}, particularly in criminology. However, if any of these steps fail---such as an insufficient number of matching minutiae, unclear classification, or incorrect relative positioning---the fingerprint cannot be considered identified.
From this process, various classification systems have emerged, all rooted in Vucetich’s foundational framework. These systems are adapted locally in different countries, with the most famous, specific, and widely accepted being the one proposed by Edward Henry in 1928 \cite{henry1928classification}. The main goal of this article is to build a model that mimics basic fingerprint patterns with directional profile structures, as shown in Fig.~\ref{Deltas}. The model presented on this paper reproduces special cases classified under the Chilean system, naturally forms minutiae structures as illustrated in Fig.~\ref{minutas}, and spontaneously exhibits the properties proposed by F. Galton in \cite{galton}.

The key to understanding fingerprint formation lies in the fixation of these patterns on the skin, a process that occurs in the womb around the sixth month of fetal development. These patterns remain with each individual from birth until death, and even beyond, making this characteristic crucial for criminal investigations and the identification of human remains. Although the exact mechanism behind the formation of these patterns is still unknown, research efforts continue. Authors such as Fleury \cite{fleury2002morphogenesis} and Glover and his team \cite{glover2023developmental} have studied how cell migration and collagen fibers interact during embryonic morphogenesis, and how these interactions influence the development of various human body structures. Their findings offer a compelling explanation for the origin of fingerprint patterns.

Another well-articulated mechanism for fingerprint pattern formation is based on non-linear elastic interactions between skin layers, as explored by M. Kücken and A. C. Newell in \cites{kucken1, kucken2}. In this model, the skin behaves as non-linear, curved elastic layers under pressure, buckling out of the surface plane and giving rise to pattern formation through the release of elastic energy. This explanation involves a fourth-order partial differential equation system with periodic boundary conditions, which yields highly accurate results. However, it falls short in generating short-range structures such as hooks, islands, or dots---like those shown in Fig.~\ref{minutas}---because the buckling process tends to produce more continuous formations. When comparing the boundary conditions with actual fingerprints, it becomes evident that fingerprint ridges do not encircle the fingertips. Instead, the ridge joints maintain a specific angle relative to the folds of the finger and the nail bed.

To address the limitations of short-range structures and boundary conditions in the Kücken model, we propose a different approach that is the main novelty of the present work: the natural curvature of the fingertip, along with the emergence of delta and center structures, guides the interaction within a two-species reaction-diffusion system. This model is based on differences in chemical composition that may lead to the formation of pores atop the ridges during the development of the ectodermal tissue layer. Reaction-diffusion systems are widely used in biology, following the seminal work of A. Turing \cite{Turing1} on the instability of homogeneous stationary solutions.

To construct the model for pattern formation, our main purpose is to establish that the equation system must be capable of generating the various types of minutiae used in human identification. Based on the various aspects and characteristics mentioned above, the model built is based on a two species reaction-diffusion PDE model, where the Schnakenberg interaction model was the one selected for this purpose because it shows sets of parameters where bifurcations and ridge endings appears naturally according to  Zhao, Zhang, and Zhu on \cite{zebrafish}. Once the model was defined, its parameters were tuned so that the typical distance between maximum on the pattern mimic the actual distances between fingerprint ridges. This was achieved using estimations from perturbation theory applied to the model, similar to those used in \cites{Meinhardt, Turing1, murray}, to interpret the system as exhibiting Turing instability, an approach also proposed by Glover and his team in \cite{glover2023developmental}. Most of the effort put on to mimic real fingerprints have had great results on the self-organization of the patterns as is shown on Cappelli and Maio's work \cite{capelli} with SFinGe, which has been the basis for other software and the complementary use of artificial intelligence on Mistry's work developing on PrintsGAN \cite{mistry2020fingerprint}, which inherits the lack in development of short range structures because of the synthetic ridge growing process, and is still limited on simulation on the basic classification of fingerprints presented by J. Vucetich. The model presented on this paper exhibits a natural finishing time for the simulation and a natural way of formation of short-range structure, besides giving a wider range of simulation on a more various kinds of fingerprint, which typically are classified as anomalies.

\section{The model}
\label{sec:Modelo}
Reaction-diffusion models have been widely studied in biological contexts \cites{Meinhardt, zebrafish, murray} due to their role in pattern formation and the wide variety of scenarios in which they can be applied. A key feature is the Turing instability \cite{Turing1}, originally introduced by Alan Turing in 1956 to explain the emergence of patterns and intricate structures in animal pelage. This instability gives rise to different structures depending on the parameters used, ranging from island-like structures in 2D domains or spike-like structures in 1D, as studied by Winter and Wei \cite{wei2013mathematical}, to more complex labyrinthine patterns in 2D, as reported by Zhao, Zhang, and Zhu \cite{zebrafish}.

The convergence properties of stationary solutions and their limits have been investigated by Takagi \cites{Takagi_condens, Takagi_stabilidad, Takagi_estimacion} in the context of Gierer--Meinhardt models, and by Masuda and Takahashi \cite{Masuda-Takahashi} in relation to the formation of biological patterns. Additionally, the existence of solutions has been explored by Antwi-Fordjour and Nkashama \cite{Unica}, contributing to an extensive theoretical framework and a broad set of mathematical tools for understanding this phenomenon. Significant progress has been made in studying the existence and regularity of spike-like structures in the one-dimensional Gierer--Meinhardt model, notably by Wei and Winter in \cite{wei2013mathematical}, and the instabilities shown on reaction-diffusion models over static and growing domains as the one studied by Castillo, Sánchez-Garduño, and Padilla in \cite{CastilloSanchezPadilla}. 

As discussed in the introduction, fingerprint minutiae can be categorized into three basic structures: terminations, bifurcations, and islands. It has been noted that cross-diffusion applied to the Schnakenberg model of two species (activator and inhibitor), under periodic or Neumann boundary conditions, can produce labyrinthine patterns that naturally form bifurcation and termination structures. Within certain parameter ranges, this also allows for the emergence of island-like structures. The model features third-degree nonlinearity and differs from other reaction-diffusion systems, such as the one studied by Wei and Winter in \cite{wei2013mathematical}, which involves a second-degree interaction in one species and a rational-type interaction in the other, posing analytical challenges for small inhibitor values.

An example of this type of behavior, observed in biology and modeled by a reaction–diffusion system, is presented by Zhao, Zhang, and Zhu in \cite{zebrafish}. In their study, modifying one of the parameters leads to different pigment pattern formations on the skin of zebrafish, each exhibiting distinct structural characteristics. This aligns with the goal of identifying a configuration that satisfies the constructed model. We explored a wide range of parameters and identified a set capable of producing the three desired structural patterns.

\medskip
To address the problem, we first define a two-dimensional domain $\Omega$ that resembles the shape of a fingertip. Since the working domain lies in the plane $\mathbb{R}^{2}$, the background curvature of the ridge pattern must reflect the fact that the ridges primarily grow in the horizontal direction. To mimic the curvature of the fingertip, the center (or nucleus) of the domain is vertically deflected. These features are encoded through a direction matrix $D$, defined as
\begin{equation*}
    D \colonequals R(\theta) S R^{-1}(\theta).
\end{equation*}
Here,
\begin{equation*}
 R(\theta) =  \begin{pmatrix}
    \cos{(\theta)} & -\sin{(\theta)}\\
    \sin{(\theta)} & \cos{(\theta)}
    \end{pmatrix} 
\end{equation*}
is a rotation matrix that determines the growth direction of the fingerprint-like pattern. This direction depends on the presence and position of deltas, centers, and the background curvature of the fingertip, making $\theta$ the parameter that encodes the tangential direction of the pattern. The matrix $S$ is a diagonal, positive-definite matrix. To ensure uniform ellipticity of the simulated model, the condition $\mathrm{det}(S) = 1$ is imposed, giving rise to a family of matrices of the form
\begin{equation*}
 S =  \begin{pmatrix}
    \gamma & 0\\
    0 & 1/\gamma
    \end{pmatrix},
\end{equation*}
where $\gamma > 1$ is a constant that ensures the diffusion is anisotropic. To mimic a more realistic fingerprint pattern, $\gamma$ represents the typical length of hooks and islands. Based on real fingerprint data, we observed that the typical ratio between the width and length of islands and hooks satisfies
\begin{equation*}
\frac{\mathrm{width}}{\mathrm{length}} = \frac{\gamma}{\dfrac{1}{\gamma}} = \gamma^{2} \in [2.5, 3.5].
\end{equation*}

As the rotation matrix $R(\theta)$ must implicitly encode the positions of the centers and deltas in the fingerprint, since they influence the angle of the ridges, the model has the potential to simulate a wide variety of patterns. To construct a complete directional map for fingerprint-like patterns, it is necessary to understand how ridge orientation changes depending on the positions of deltas and centers. 

Inspired by the directional maps studied by Cappelli and Maio in~\cite{capelli} for the development of a synthetic fingerprint generator, we propose that the directional angle of ridge growth can be obtained by knowing the positions of centers ($\vec{x_{C}}$) and deltas ($\vec{x_{D}}$), via
\begin{equation}
    \theta (\vec{x}) = \theta_{0} (\vec{x}) + \alpha \sum_{C=1}^{N_{C}}\arctan(\vec{x_{C}} - \vec{x}) + \beta \sum_{D=1}^{N_{D}}\arctan(\vec{x_{D}} - \vec{x}).
    \label{Ec:Cappelli1}
\end{equation}
Here, $\theta_{0}$ is a parameter that represents a background flexion or twist independent of the centers or deltas. The coefficient $\alpha$ is the Poincaré index associated with the centers, and $\beta$ is the Poincaré index associated with the deltas, which has previously been studied only in the context of directional maps~\cites{sherlock1993model,patmasari2022determination}.

\begin{figure}[ht]
    \centering
    \includegraphics[width=0.4\textwidth]{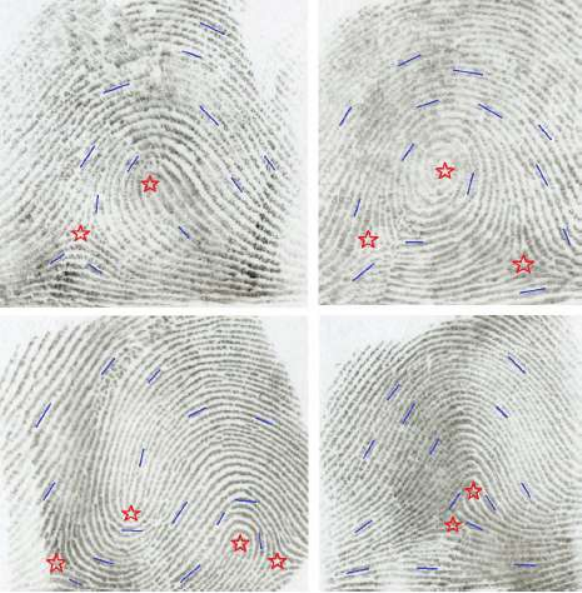}
    \caption{A set of fingerprints used to obtain the values in Table~\ref{Tabla}. Red stars indicate the positions of centers and deltas, while blue segments are used to measure ridge angles.}
    \label{angulos}
\end{figure}

To obtain the parameters $\theta_{0}$, $\alpha$, and $\beta$ in a realistic way, we performed a least-squares minimization between equation~\eqref{Ec:Cappelli1} and empirical data extracted from real fingerprints, such as those shown in Fig.~\ref{angulos}. A dataset of 200 fingerprints was collected from 20 consenting individuals. Each fingerprint was annotated with the positions of its centers and deltas, and approximately 15 ridge orientation samples were recorded per domain. The fingerprints used to construct Table~\ref{Tabla} were categorized according to the following classification: 21 arches (including tented arches), 55 radial loops, 61 ulnar loops, 36 mononuclear whorls, and 27 binuclear whorls.

\begin{table}[ht]
\centering
\begin{tabular}{ |c|c|c|c|c|  }
 \hline
 Parameters (rad) & Pinned  Arches & Loops & Mononuclear Whorls & Binuclear Whorls\\
 \hline
 $\theta_{0}$   & 0.78 $\pm$ 0.21    & 0.61 $\pm$ 0.24 &   0.91 $\pm$ 0.14 & 0.98 $\pm$ 0.75 \\
 $\alpha$ &   -0.45 $\pm$ 0.25  & -0.51 $\pm$ 0.12   & -1.02 $\pm$ 0.07 & -0.48 $\pm$ 0.18 \\
 $\beta$ & 0.52 $\pm$ 0.012 & 0.49 $\pm$ 0.018 &  0.50 $\pm$ 0.017 & 0.52 $\pm$ 0.22 \\
 \hline
\end{tabular} 
\vspace*{5pt}
\caption{Parameter values obtained through a least-squares minimization process.}
\label{Tabla}
\end{table}

The statistical least-squares minimization process shows strong agreement between the observed ridge behavior and the results reported by Cappelli, whose simulations used $\alpha = -\frac{1}{2}$ and $\beta = \frac{1}{2}$ as input parameters. Table~\ref{Tabla} suggests that mononuclear whorls are consistently composed of two closely positioned centers (i.e., $N_{C} = N_{D}$), and that the average Poincaré index of fingerprint patterns is zero.

\medskip 
To define appropriate boundary conditions, we first examined how ridge patterns behave near the fingernail and along the lateral edges of the fingertip in real fingerprints, including arches, loops, and whorls. Near the borders, ridge orientation tends to align horizontally, following the tangential direction modeled by the matrix $D$. This observation motivated the use of a Neumann-like boundary condition, expressed as $D \nabla u \cdot \nu = 0 $ on $\partial \Omega$, to better replicate realistic fingerprint patterns.

Finally, we propose the following model as a candidate for simulating fingerprint-like pattern formation:
\begin{equation}\label{modelo}
\left\{
\begin{array}{rcll}
     \partial_{t} u &=& D_{uu}\nabla \cdot (D \nabla u) + D_{uv} \triangle v + K(a - u + u^{2} v)&\mbox{in } \Omega\\
     \partial_{t} v &=& D_{vv}\nabla \cdot (D \nabla v) + D_{vu} \triangle u + K(b - c v - u^{2} v)&\mbox{in } \Omega\\
     D\nabla u \cdot \hat{n} &=& 0 &\mbox{on } \partial \Omega\\
     D\nabla v \cdot \hat{n} &=& 0&\mbox{on } \partial \Omega\\
     u(0,x,y) &=& u_{0}(x,y)&\mbox{for }(x,y)\in \Omega \\
     v(0,x,y) &=& v_{0}(x,y)&\mbox{for }(x,y)\in \Omega.
\end{array}
\right. 
\end{equation}
Here, the initial conditions $u_0, v_0 \in L^2(\Omega)$, and the parameters $D_{uu}, D_{uv}, D_{vu}, D_{vv}, K, a, b,$ and $c$ are positive constants, with
$$
D_{uu} D_{vv} - D_{uv} D_{vu} > 0.
$$
Note that the above condition guarantees that the linearized PDE system around the unique homogeneous stationary solution at $c = 0$, for the isotropic diffusion case, corresponds to an elliptic variational problem, making the PDE problem suitable for numerical simulations via functional minimization. In this case, the homogeneous stationary solution for $c=0$ is uniquely given by $u = a + b$ and $v = \dfrac{b}{(a + b)^2}$, and the initial conditions are modeled as stochastic perturbations around it.

To develop a model capable of generating not only the basic fingerprint types described by previous authors, but also a broader range of fingerprint-like patterns, we incorporated the Chilean fingerprint classification key to define the range of the variable $\theta$.

In 1929, Chilean police were tasked with studying and adapting the Argentinian fingerprint classification system. Today, the Chilean system comprises fourteen distinct classification values, with an additional category used in cases of amputation. This framework is rooted in the principles established by Vucetich and Henry~\cites{vucetich1896instrucciones,vucetich1904dactiloscopia,henry1928classification}, which define various fingerprint characteristics. These principles are further elaborated in Alberto Teke Schlicht’s book~\cite{teke2010medicina}, a reference still widely used for human identification and forensic medicine. The Chilean fingerprint classification key is presented in Table~\ref{Chilena}.

\begin{table}[ht]
\centering
\begin{tabular}{|@{\hspace{5pt}}l@{\hspace{5pt}}|@{\hspace{5pt}}l@{\hspace{5pt}}|@{\hspace{5pt}}l@{\hspace{5pt}}|}
\hline
\textbf{0:} Simple arch & \textbf{5:} External variable loop & \textbf{a:} Binuclear medial whorl \\
\textbf{1:} Tented arch & \textbf{6:} Mononuclear internal whorl & \textbf{b:} Binuclear external whorl \\
\textbf{2:} Simple internal loop & \textbf{7:} Mononuclear medial whorl & \textbf{c:} Hooked whorl or hooked loop \\
\textbf{3:} Internal variable loop & \textbf{8:} Mononuclear external whorl & \textbf{x:} Unclassifiable \\
\textbf{4:} Simple external loop & \textbf{9:} Binuclear internal whorl & \textbf{z:} Amputation \\
\hline
\end{tabular}
\vspace*{5pt}
\caption{Fingerprint classification used in the Chilean system.}
\label{Chilena}
\end{table}

This classification key has proven highly effective in forensic work conducted by Chilean police, as it enables the identification of altered or miscategorized fingerprints. This is particularly useful in cases where chirality-related anomalies, frequently observed in day-to-day forensic analysis, affect pattern interpretation.

Value 0 corresponds to a normal arch, which presents neither deltas nor centers. Value 1 denotes an arch with a center, although the ridges do not encircle it; instead, they pass above and below. Values 2 and 4 represent loops with fixed delta positions that are independent of the hand being analyzed. On the right hand, value 2 indicates a radial loop, while value 4 indicates an ulnar loop; for the left hand, these interpretations are reversed. Values 3 and 5 also correspond to loops with delta positions matching those of values 2 and 4, respectively, but they exhibit atypical ridge flow around the center that does not conform to the delta’s position.

Values 6--9, as well as a and b, distinguish different types of whorls based on how the delta branches spread within the pattern and the number of identifiable centers, depending on which lower branch overlaps the other. Value c refers to loops or whorls characterized by a pseudo-delta appearing in the central region of the fingerprint, an anomaly noted in Henry’s classification system. Value x is assigned to fingerprints that are too damaged to classify or contain scars that significantly distort the pattern. Value z is used when the finger has been amputated.

\smallskip 
To distinguish between mononuclear and binuclear whorls, one must examine the extensions of the guideline ridges that emerge from the deltas. Both centers and deltas correspond to singular points within the fingerprint pattern: centers are associated with a single guideline ridge, while deltas typically have three. Of these, one ridge usually points toward the exterior of the domain, while the other two are used for classification within the Chilean system.

For horizontally oriented guideline ridges, whorl classification depends on how the ridge from the left delta interacts with the right delta. If it extends and passes above three ridges over the right delta, the whorl is categorized as internal. If it passes below three ridges under the right delta, it is categorized as external. If neither condition is met, the whorl is classified as medial.

For vertically oriented guideline ridges, classification depends on their interaction with the centers. If one of these ridges extends and separates the centers, the pattern is classified as a binuclear whorl. If both ridges curve around the centers without separating them, the pattern is classified as mononuclear.

Based on these structural details, we developed an open-source software named GenCHSin (Chilean Synthetic Fingerprint Generator), which can generate a wide range of fingerprint patterns using the Chilean classification key presented in Table~\ref{Chilena}.

\section{Numerical simulations}
There have been various approaches to generating synthetic fingerprints. One notable example is the SFinGe program developed by Cappelli, Maio, and their team~\cite{capelli}. This program, based on directional maps and seeded growth, generates synthetic fingerprints through wave propagation and the addition of noise. The results are generally very good, although they depend heavily on the wavelength used.

The SFinGe algorithm has served as a foundation for several subsequent developments: Priesnitz and his team created SynCoLFinGer~\cite{priesnitz2022syncolfinger}; Mistry’s group used it to develop a fingerprint search algorithm for training machine learning models~\cite{mistry2020fingerprint}; Engelsma’s group adopted its ideas to build PrintsGAN~\cite{engelsma2022printsgan}; and Wyzykowski’s team extended the approach to incorporate a third recognition level for detecting and characterizing pores~\cite{wyzykowski2021level}.

\begin{figure}[ht]
    \centering
    \includegraphics[width=0.6\textwidth]{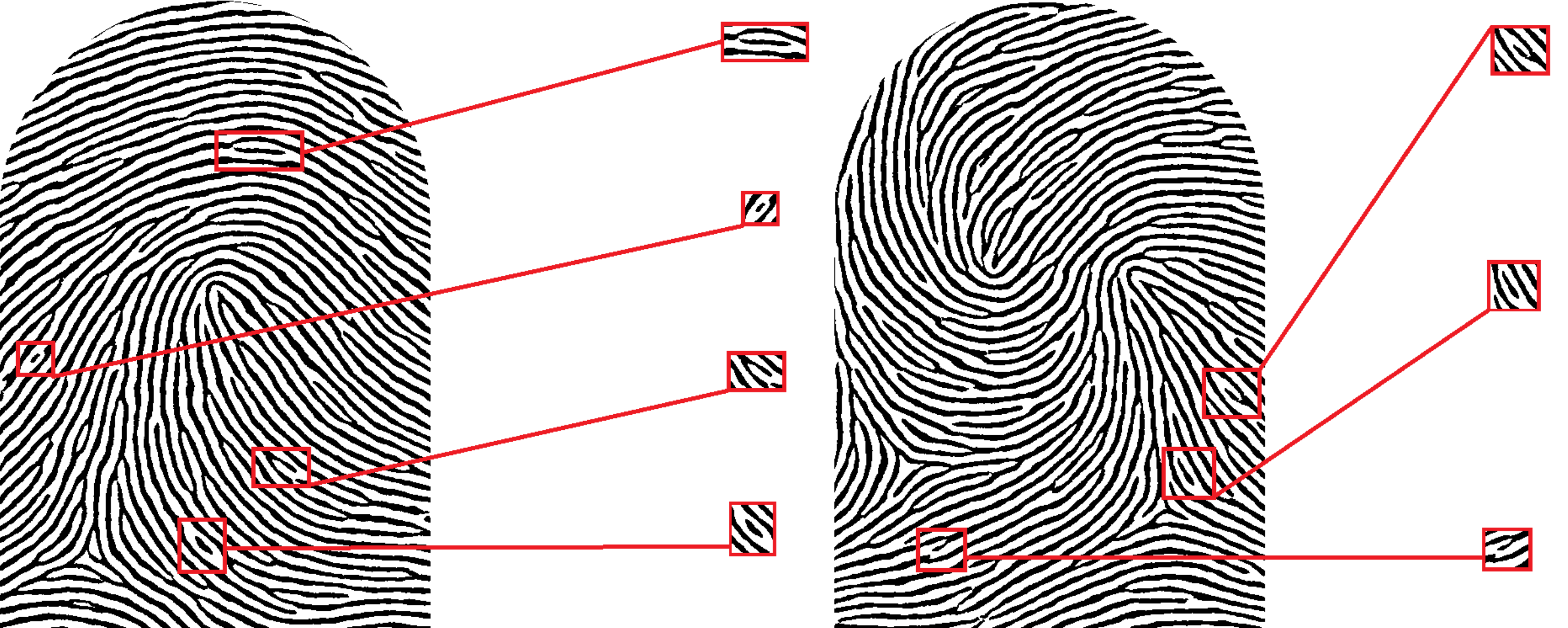}
    \caption{Examples of two distinct synthetic fingerprints generated using the proposed model. Red rectangles highlight the presence of short-range structures such as hooks, islands, and dots in each image.}
    \label{corto_alcance}
\end{figure}

While these programs produce synthetic fingerprints that closely resemble real ones, they often fail to reproduce short-range structures such as islands, dots, or hooks, an inherent limitation of the Cappelli and Maio method. As shown in Fig.~\ref{corto_alcance}, such fine-scale features emerge naturally in the model presented in Section~\ref{sec:Modelo}.

To obtain numerical results from the proposed model, we developed the open-source software GenCHSin. The implementation began in Python, incorporating the linear solver from the FEniCS Project, an open-source platform for solving partial differential equations using Python and C++. FEniCS is capable of simulating a variety of elliptic problems, including the heat equation, Navier--Stokes equation, and Laplace equation.

This solver was used to simulate the model presented in~\eqref{modelo}, employing finite elements for spatial discretization and finite differences for time integration. To reduce the simulation time, we adopted an explicit resolution scheme, which is shown as follows:
\begin{equation}\label{esquema}
\left\{
\begin{array}{rcll}
     \dfrac{u^{n+1}-u^{n}}{\Delta t} &=& \nabla \cdot (D \nabla u^{n+1}) + D_{uv} \triangle v^{n} + K(a - u^{n+1} + u^{n+1}u^{n}v^{n}) &\mbox{in } \Omega\\[10pt]  
    \dfrac{v^{n+1}-v^{n}}{\Delta t} &=& \nabla \cdot (D \nabla v^{n+1}) + D_{vu} \triangle u^{n} + K(b - cv^{n+1} - u^{n}u^{n}v^{n+1}) &\mbox{in } \Omega.
\end{array}
\right. 
\end{equation}
Each of the equations in \eqref{esquema} gives rise to minimization functionals 
$$J_{1}(u^{n+1},u^{n},v^{n})\quad \mbox{and}\quad J_{2}(v^{n+1},u^{n},v^{n}),$$
constructed inside the code by integrating the system over the domain and applying the boundary conditions. These functionals are then used by the FEniCS linear solver (see for instance \cites{langtangen2017solving, allaire2005analyse, aziz2014mathematical}), integrated into GenCHSin, to build the numerical solutions. We recall that, to ensure a wide variety of patterns, the initial conditions consisted of uniform random perturbations around the homogeneous stationary solution of the system when $c = 0$, as described in Section~\ref{sec:Modelo}.

Upon completing 4000 iterations, the simulation process concludes and the program saves an image of the final state of the activator $u$, representing a stationary numerical solution of the model described in~\eqref{modelo} and previously introduced in Section~\ref{sec:Modelo}. The simulation terminates when the minimization functionals  $J_{1}(u^{n+1},u^{n},v^{n})$ and  $J_{2}(v^{n+1},u^{n},v^{n})$, derived from the variational formulation of the linearized problem, can no longer be improved. The number of iterations was determined empirically through trial and error based on initial realizations.

To ensure that the simulated patterns resembled the shape of a real fingerprint on a flat 2D surface, the domain $\Omega$ was defined as a square of 12 units in length with a semicircle of 6 units in radius attached to its top. For the numerical simulation space $H^{1}(\Omega) \times H^{1}(\Omega)$, we used 200 nodes based on second-order Lagrange polynomials to ensure smoothness and continuity across the domain. The number of nodes was adjusted to obtain sufficiently regular solutions without compromising computational efficiency. For a detailed discussion of Sobolev spaces and their role in partial differential equations, we refer the reader to the classical book \cite{evans2022partial}.

\begin{figure}[ht]
    \centering
    \includegraphics[width=0.4\textwidth]{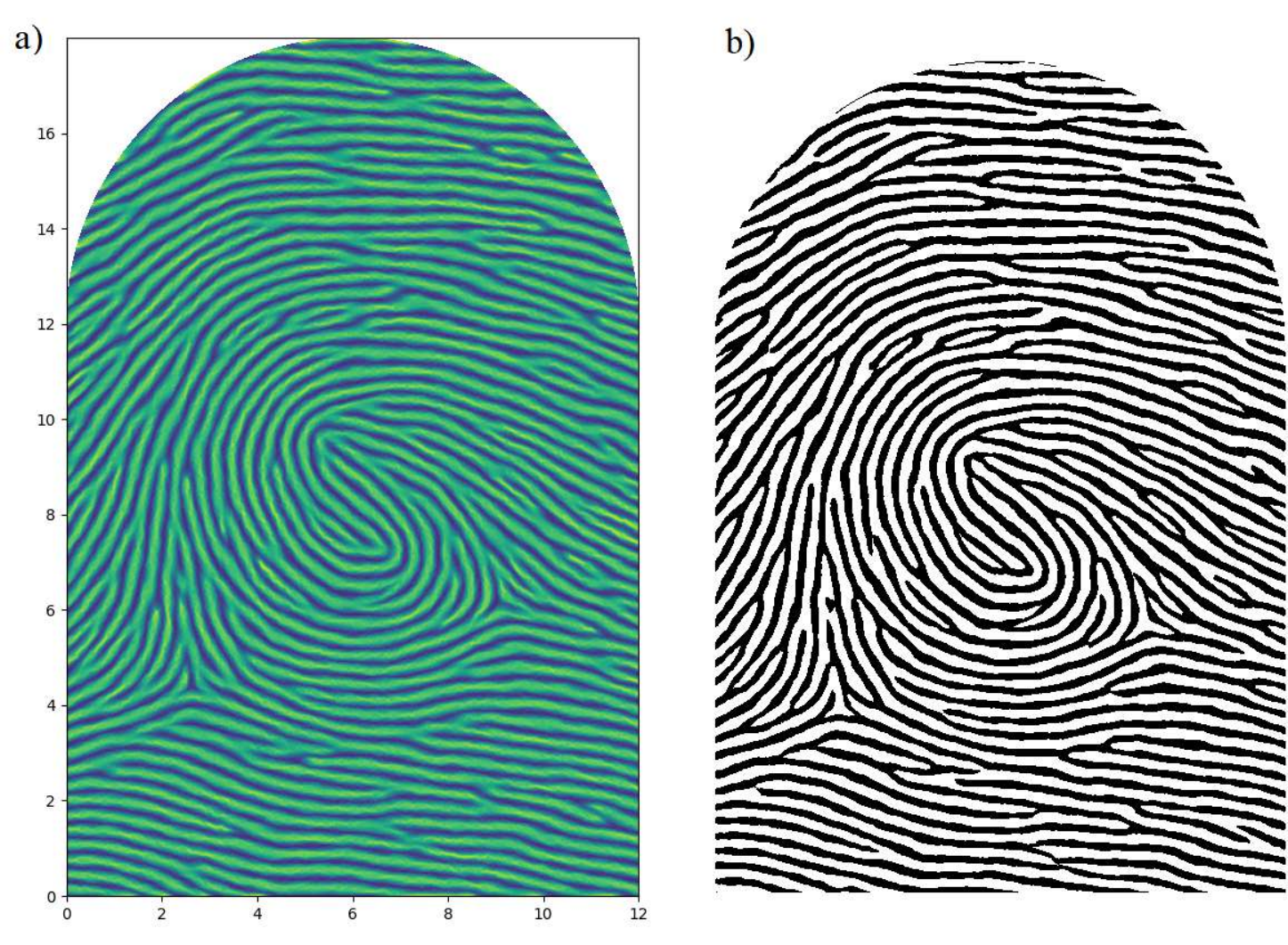}
    \caption{a) Plot of the activator $u$ at the end of the simulation process. b) Result of applying the cutting and black-and-white filter to the corresponding image.}
    \label{color_vs_bw}
\end{figure}

To make the simulated patterns more visually comparable to real fingerprints, the generated images are processed using a black-and-white filter as shown in Fig. \ref{color_vs_bw}. This filter retains only the regions where the solution exceeds a certain threshold, mimicking the appearance of an ink-based fingerprint impression. This post-processing step allows for a qualitative analysis of the structures formed during the simulation.

The threshold for the black-and-white filter is set at 120 on a scale from 0 to 255 (the standard grayscale range), chosen so that the characteristic width of the lines closely resembles that of real fingerprint impressions. After applying the filter, the images are cropped to retain only the domain $\Omega$. These processed images are then analyzed to compare identification-relevant features and behavioral patterns with those observed in real fingerprints.

\medskip 
To contextualize our approach, it is useful to examine the state of the art in synthetic fingerprint generation models, such as the work of Cappelli (see Section~\ref{sec:Modelo}) and Kücken (see Section~\ref{sec:Intro}). The model introduced by Cappelli~\cite{capelli} is based on wave-like evolution, simulating ridge growth from multiple seed points. It successfully reproduces various fingerprint minutiae, although the termination of the simulation is manually controlled by the user.

In contrast, Kücken’s model~\cites{kucken1,kucken2} is formulated as a fourth-order non-linear energy minimization system, offering strong theoretical foundations and compelling results. However, it does not capture fine-scale structures such as points, hooks, and islands, as illustrated in Fig.~\ref{minutas}. Additionally, the model employs periodic boundary conditions, which are mathematically convenient but lack the realism needed to reflect the natural boundaries of actual fingerprints.

Both models are limited in their ability to reproduce the full diversity of whorls and loops found in human fingerprints, particularly those classified by Henry as anomalies. Our model aims to address these limitations by incorporating structural variability and finer morphological details.

\begin{figure}[ht]
    \centering
    \includegraphics[width=0.8\textwidth]{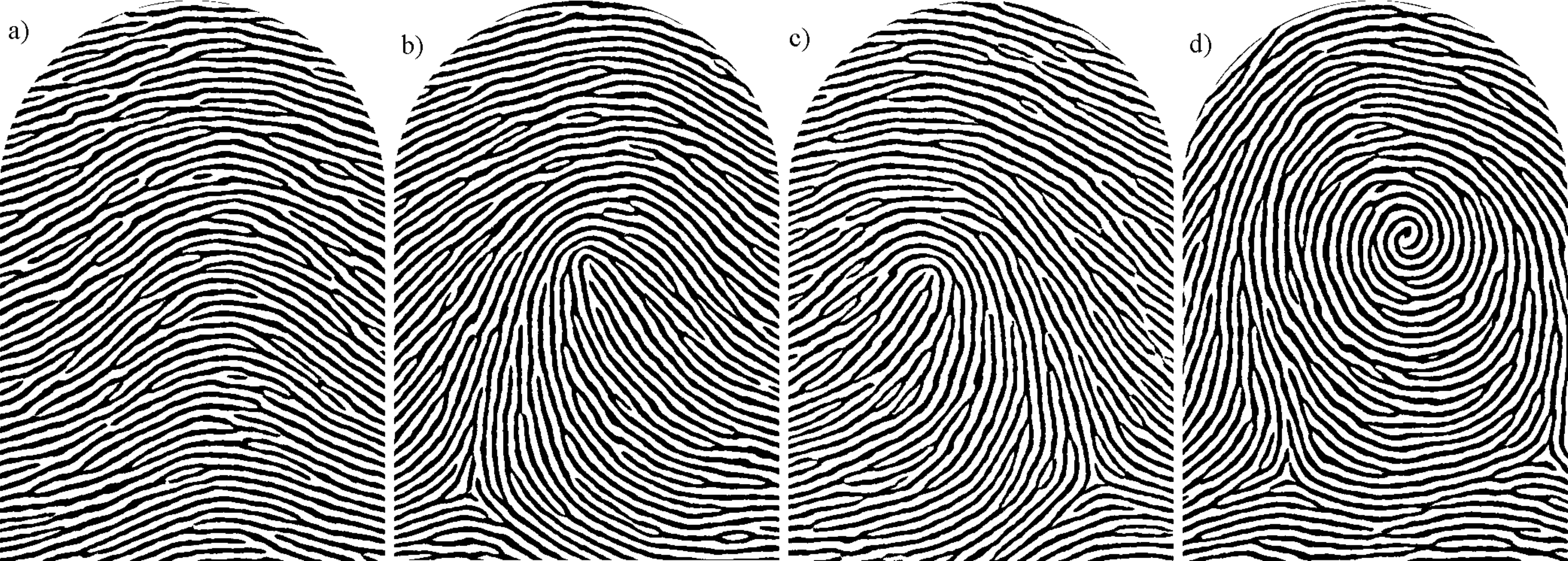}
    \caption{Different results from the simulation of the proposed model correspond to the fundamental fingerprint classifications, including:  
    a) Simple arch, b) Simple external loop, c) Simple internal loop, and d) Mononuclear medial whorl.}
    \label{huellitas}
\end{figure}

Building on the comparison with existing models, the approach presented in this paper offers several advantages. It exhibits stationary states, with simulations converging through finite-element energy minimization. The model is capable of generating all minutiae-related structures shown in Fig.~\ref{minutas}, and can reproduce the main fingerprint types---arch, loop, and whorl---according to Vucetich’s fundamental classification. The presence of stationary states provides a natural stopping criterion for the simulation. As shown in Fig. \ref{huellitas}, the model presented on this paper is capable of generate the fingerprint-like patterns that are in good agreement with the state-of-art models mentioned in Section \ref{sec:Intro}, preserving the shape and auto-organization expected on real fingerprints, exhibiting the wide range of minutiae expected to find.

\begin{figure}[ht]
    \centering
    \includegraphics[width=0.95\textwidth]{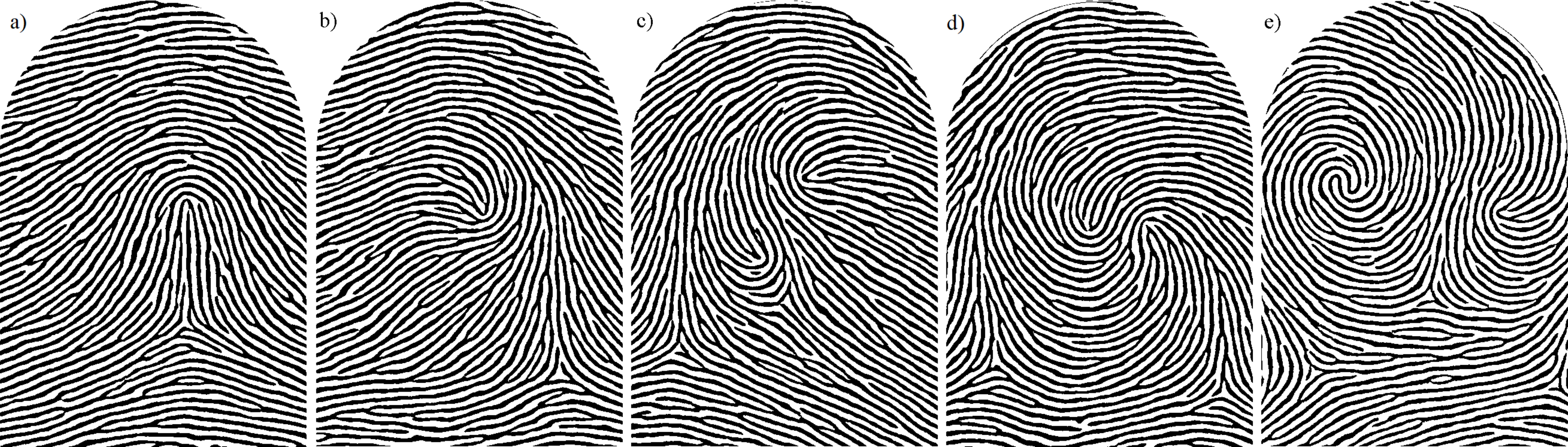}
    \caption{Different simulation results of the proposed model, corresponding to the Chilean fingerprint classification key: a) Tented arch, b) Internal variable loop, c) External hooked loop, d) Binuclear medial whorl, and e) Hooked whorl.}
    \label{huellitas2}
\end{figure}

\smallskip
In addition, the model can produce a wide variety of ridge configurations observed in human fingerprints as shown in Fig. \ref{huellitas2}, which in some aspects cannot be easily classified by Vucetich’s fundamental classification. These kind of patterns are not simulated or studied on the state-of-art fingerprint pattern generator models, which is the reason why it was used the Chilean classification key: to ensure the wide variety of possible patterns that appears in real fingerprints. 

Its formulation is based on second-order equations, which makes it relatively easy to implement. The different fingerprint patterns categorized by the Chilean classification key can be reproduced without difficulty, and short-range structures such as hooks, islands, and dots emerge naturally, as illustrated in Fig.~\ref{corto_alcance}.

\section{Validation of the model}
\label{sec:Validez}
Among the various fingerprint generation models studied over the years, the one proposed here exhibits key characteristics of an evolution-based approach capable of simulating fingerprint formation while retaining mathematical properties suitable for further analysis. The stationary solutions to which the simulations converge are stable under numerical perturbations, mirroring the permanence of human fingerprints over time, and thus lend credibility to the model.

The average number of minutiae found in real fingerprint patterns ranges from 60 to 130 per fingerprint. This variation depends on the number of deltas present in the pattern, with fewer deltas typically resulting in fewer minutiae. These values are consistent with the average number of minutiae observed in real fingerprints \cites{galton,gutierrez2011distribution,gutierrez2012there,stoney1986critical}. Additionally, the number of ridges crossing Galton’s line in loops and whorls, or located within the nuclear region in arches, varies between 5 and 20 depending on the fingerprint type analyzed, aligning well with previous studies~\cites{stoney1986critical,galton}.

Bridges, hooks, islands, and points are particularly used in forensic fingerprint analysis for human identification due to their rarity, making them strong indicators that the observed pattern belongs to a particular fingerprint. These types of minutiae appear naturally in the model presented, as shown in Fig.~\ref{corto_alcance}, regardless of the simulated pattern type. However, such minutiae are very rare in real fingerprints, as studied by Gutierrez--Redomero and her team in Spanish and Argentinian populations \cites{rivalderia2017study, gutierrez2011distribution, gutierrez2012there}, and tend to be absent in state-of-the-art models, as previously mentioned.

\begin{figure}[ht!]
    \centering
    \includegraphics[width=0.65\textwidth]{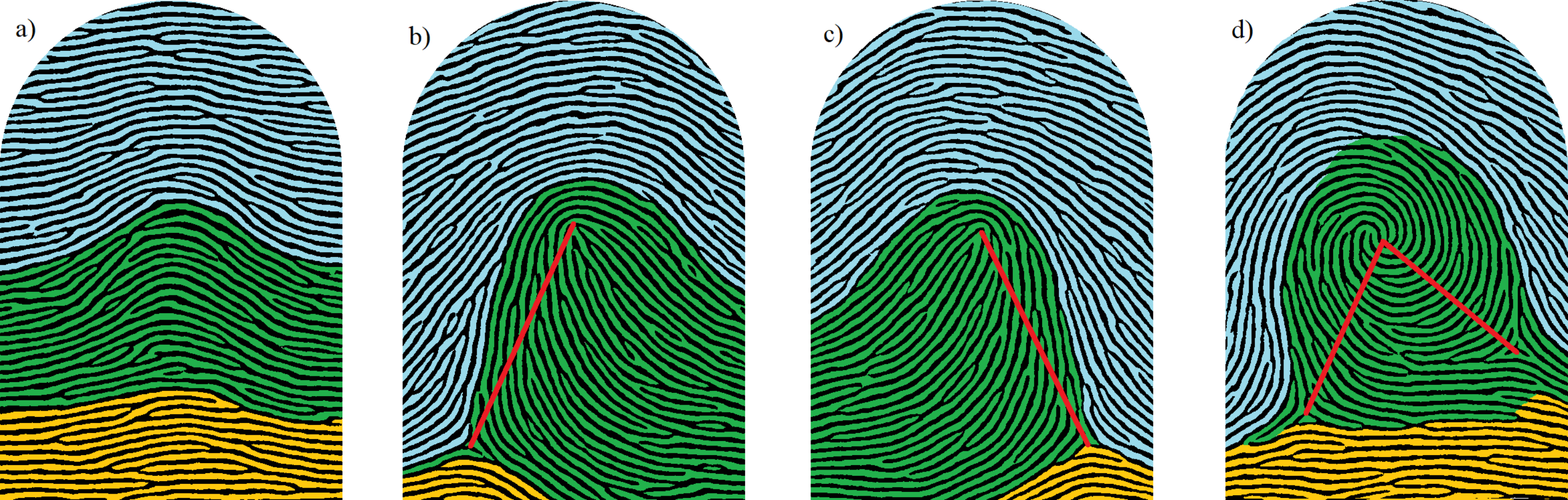}
    \includegraphics[width=0.65\textwidth]{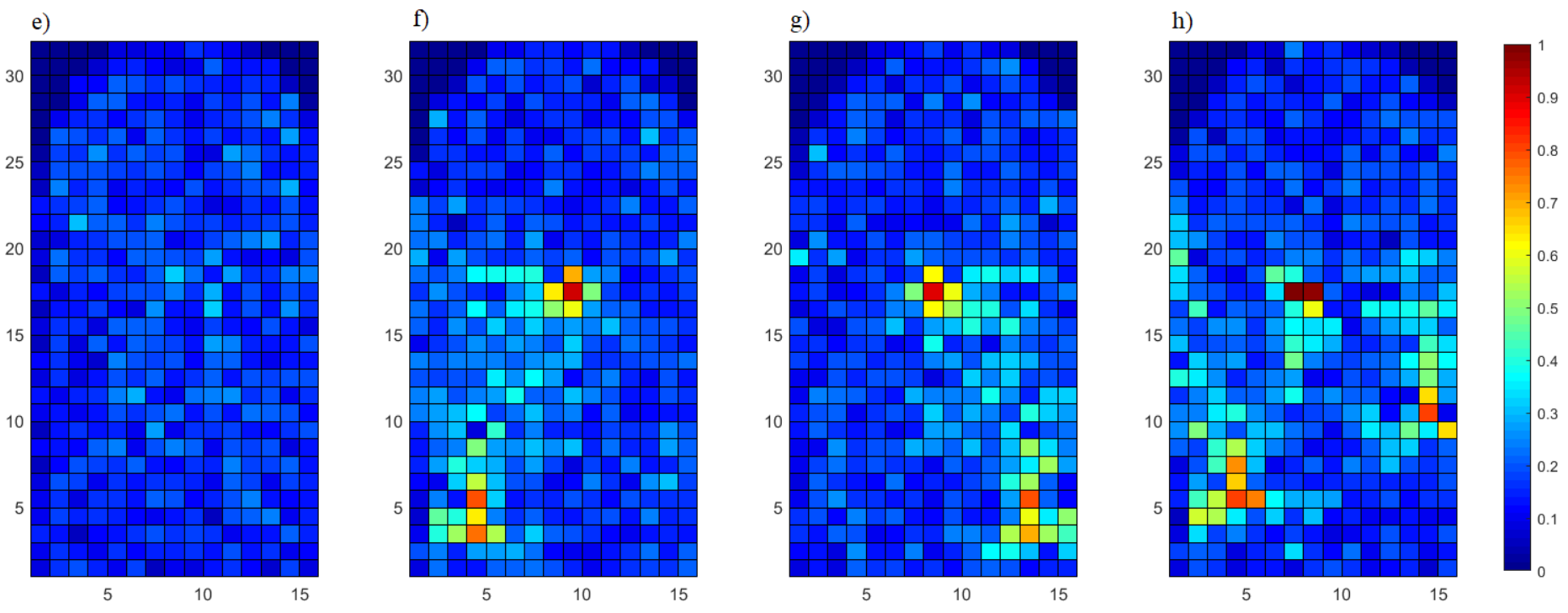}
    \includegraphics[width=0.35\textwidth]{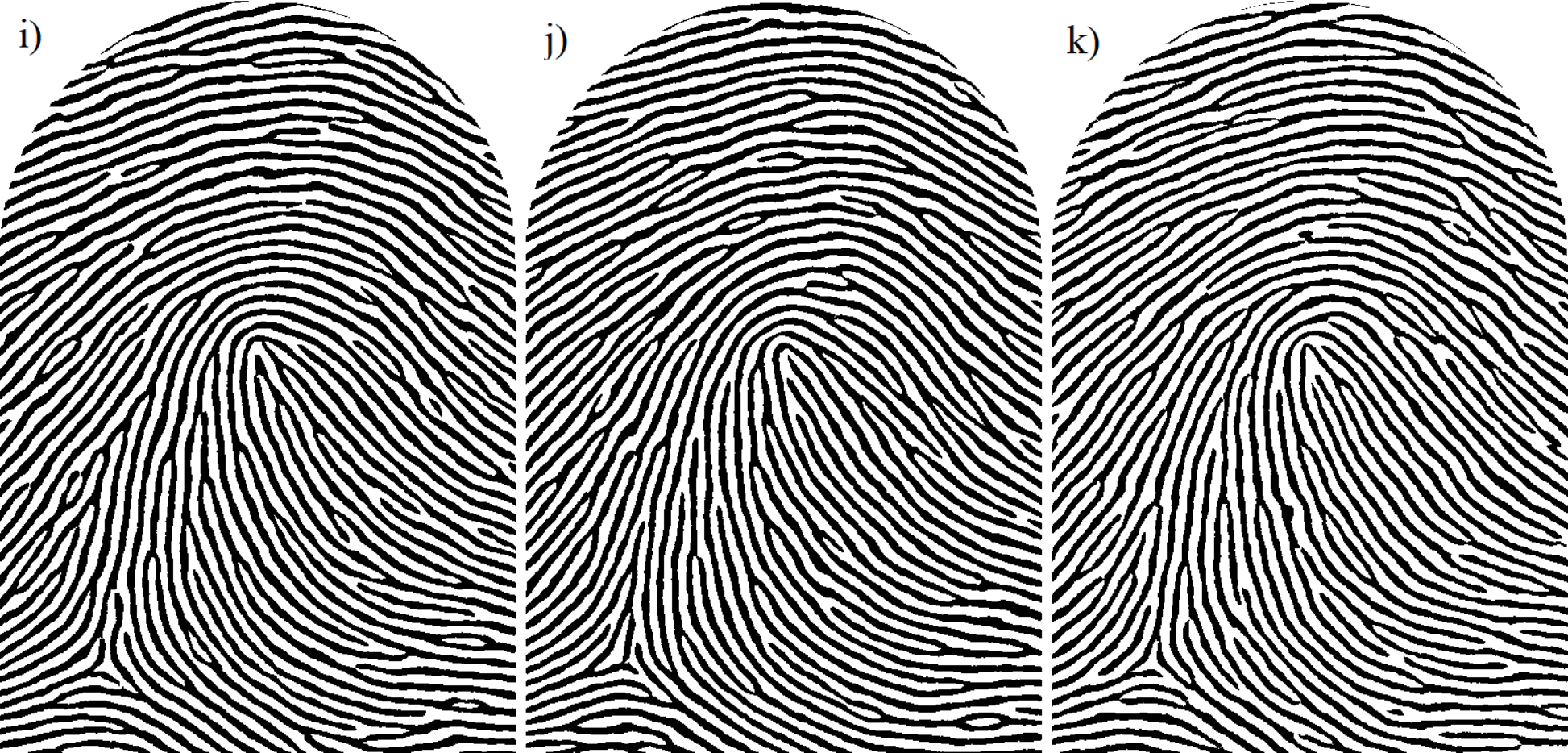}
    \caption{Examples of well-defined zones on synthetic fingerprints that are of high interest in fingerprint forensic science, where blue corresponds to marginal zones, green corresponds to nuclear zones, and yellow corresponds to basal zones. These correspond to: a) Simple arch, b) Simple external loop, c) Simple internal loop, and d) Mononuclear external whorl. The red lines drawn over the diagrams correspond to Galton's lines for each classification, which are typical characteristics used as references in fingerprint identification. The basal, nuclear, and marginal zones are divided by the ridge organization given by the position of centers and deltas, where the nuclear zone is bounded at the upper limit by the upper delta branch that extends following the directional map, and at the bottom limit by the lower delta branch that also extends along the directional map. For arches, the nuclear zone is estimated as the middle third of the domain. The lower plots correspond to the spatial distribution of minutiae appearance on each of the fingerprints shown above, corresponding to e) Simple arch, f) Simple external loop, g) Simple internal loop, and h) Mononuclear external whorl. i), j), and k) correspond to examples of simulating normal external loops with the exact same parameters $D_{uu}, D_{uv}, D_{vu}, D_{vv}, K, a, b$ and $c$, but differing only in the stochastic initial conditions, showing the kind of sets studied.}
    \label{zonas}
\end{figure}

To compare the morphological aspects \cites{lubian2010dactiloscopia, Silva, galton} and the criminological aspects of the model \cites{locard1923manuel, vucetich1896instrucciones, vucetich1904dactiloscopia, teke2010medicina} with the perspective of real fingerprints, several synthetic fingerprints were tested for the fundamental characteristics that make a pattern classifiable as a fingerprint. These include Galton's line, ridge guidelines, and the presence of basal, nuclear, and marginal zones, features illustrated in Fig.~\ref{zonas}, which allow these synthetic patterns to be categorized as plausible fingerprint structures.

For a more detailed analysis we have performed 100 different simulations on four synthetic fingerprint kinds of patterns (simple arch, simple internal loop, simple external loop and mononuclear external whorl) to compare the quantitative and qualitative properties shown by the model with real fingerprint properties. For this task, all the parameters of the model have been fixed for each of the fingerprint patterns simulated, which means that the constants $D_{uu}, D_{uv}, D_{vu}, D_{vv}, K, a, b$ and $c$ are the same for the 400 simulations, and the matrix $D$ is the same for each kind of pattern (meaning all simple arches have the same directional map, all simple internal loops have the same directional map, etc.). Each iteration is performed using different random initial conditions.

To build the spatial distribution of minutiae, the images were divided into 512 independent rectangles, each containing no more than three independent ridges, to analyze the presence or absence of minutiae according to Fig. \ref{minutas}. The procedure shown in Fig. \ref{zonas} indicates that loops and whorls tend to organize minutiae appearance along Galton's line, while arches display a more uniform distribution, as reported in \cites{galton,stoney1986critical,gutierrez2011distribution}. This has a direct impact on minutiae recognition base models, leading to the separation of marginal, nuclear, and basal zones in synthetic fingerprints that are analogous to the marginal, nuclear, and basal zones in real fingerprints, with minutiae density and distribution similar to those reported in \cites{lubian2010dactiloscopia,maltoni2009handbook,sourcebook}.

\begin{figure}[ht]
    \centering
    \includegraphics[width=0.9\textwidth]{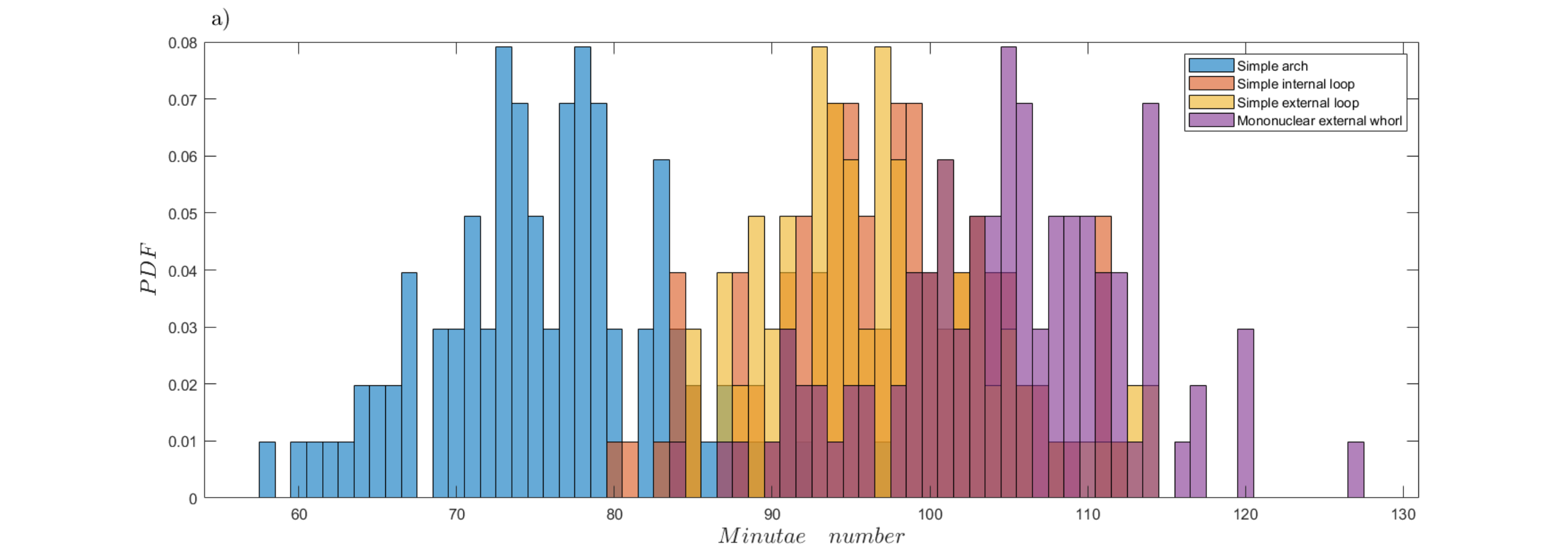}
    \includegraphics[width=0.9\textwidth]{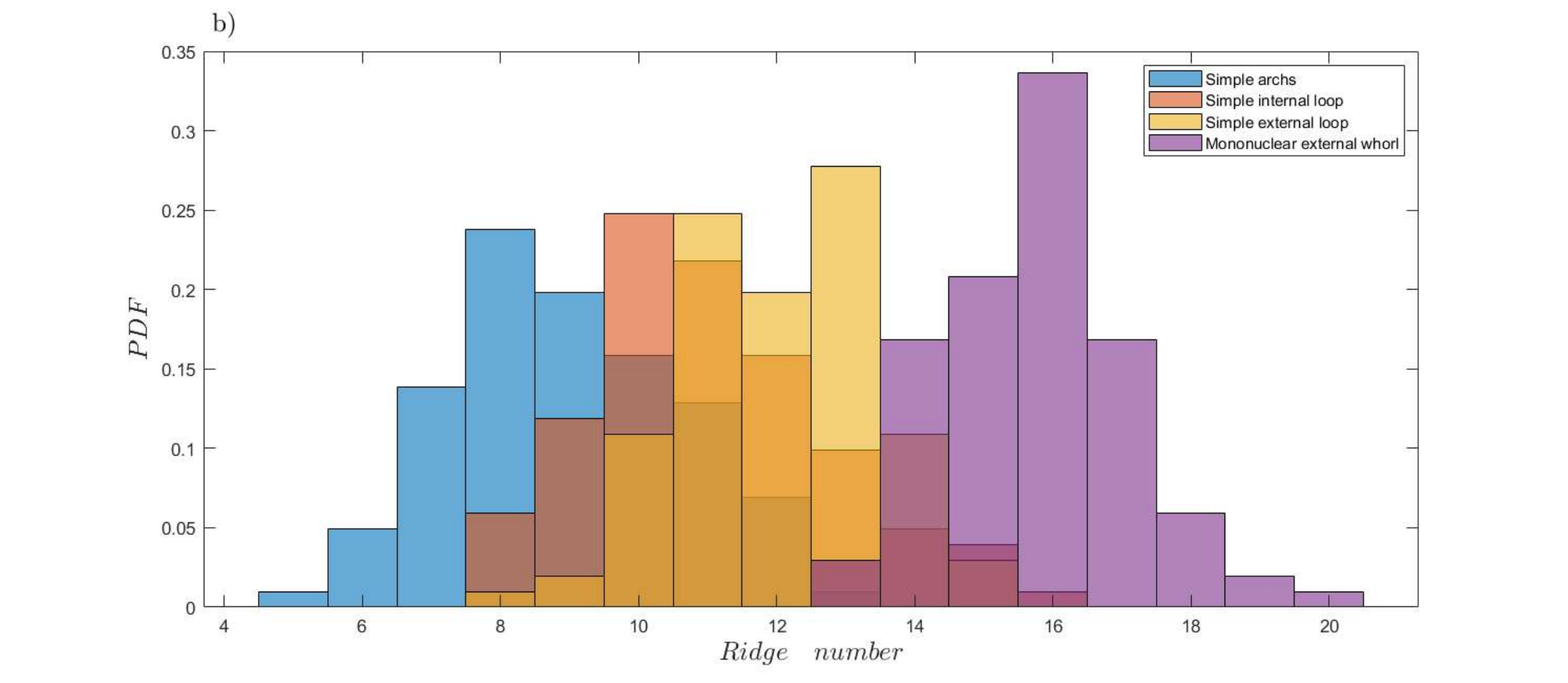}
    \caption{Plots for a) the distribution of the number of minutiae on synthetic fingerprints depending on the kind of fingerprint simulated, and b) the distribution of the number of synthetic ridges over Galton's line. For arches, as they don't have a center or delta, the number of synthetic ridges is calculated using the middle third area and counting the number of synthetic ridges along a vertical line.}
    \label{ploteos_minucias}
\end{figure}

\smallskip 
With this procedure of categorizing minutiae, it was possible to count the number of minutiae inside each domain. In Fig. \ref{ploteos_minucias}, a) shows the distribution of the number of minutiae, ranging between 60 and 130. The fewer deltas and centers a fingerprint has, the fewer minutiae tend to appear. Meanwhile, in Fig.~\ref{ploteos_minucias}, b) shows that the number of ridges over Galton's line in the case of loops and whorls, or the number of ridges in the middle third in the case of arches, tends to increase with the number of deltas. This is related to the ridge density present in each kind of fingerprint. These two indicators are in good agreement with what is found in real fingerprints according to various authors \cites{galton,gutierrez2011distribution,gutierrez2012there,stoney1986critical}.

For additional verification, several synthetic patterns generated by the model were tested using the AFIS (Automated Fingerprint Identification System) computational system \cites{sourcebook, jain2001automated, maltoni2017automated,egli2007evidence}, software commonly used to compare fingerprints obtained by forensic experts with an internal database. This system relies on the position and orientation of ridges to suggest possible candidates for further comparison. AFIS is widely used by law enforcement agencies around the world, including the Chilean police, for biometric identification tasks.

\begin{figure}[ht]
    \centering
    \includegraphics[width=0.55\textwidth]{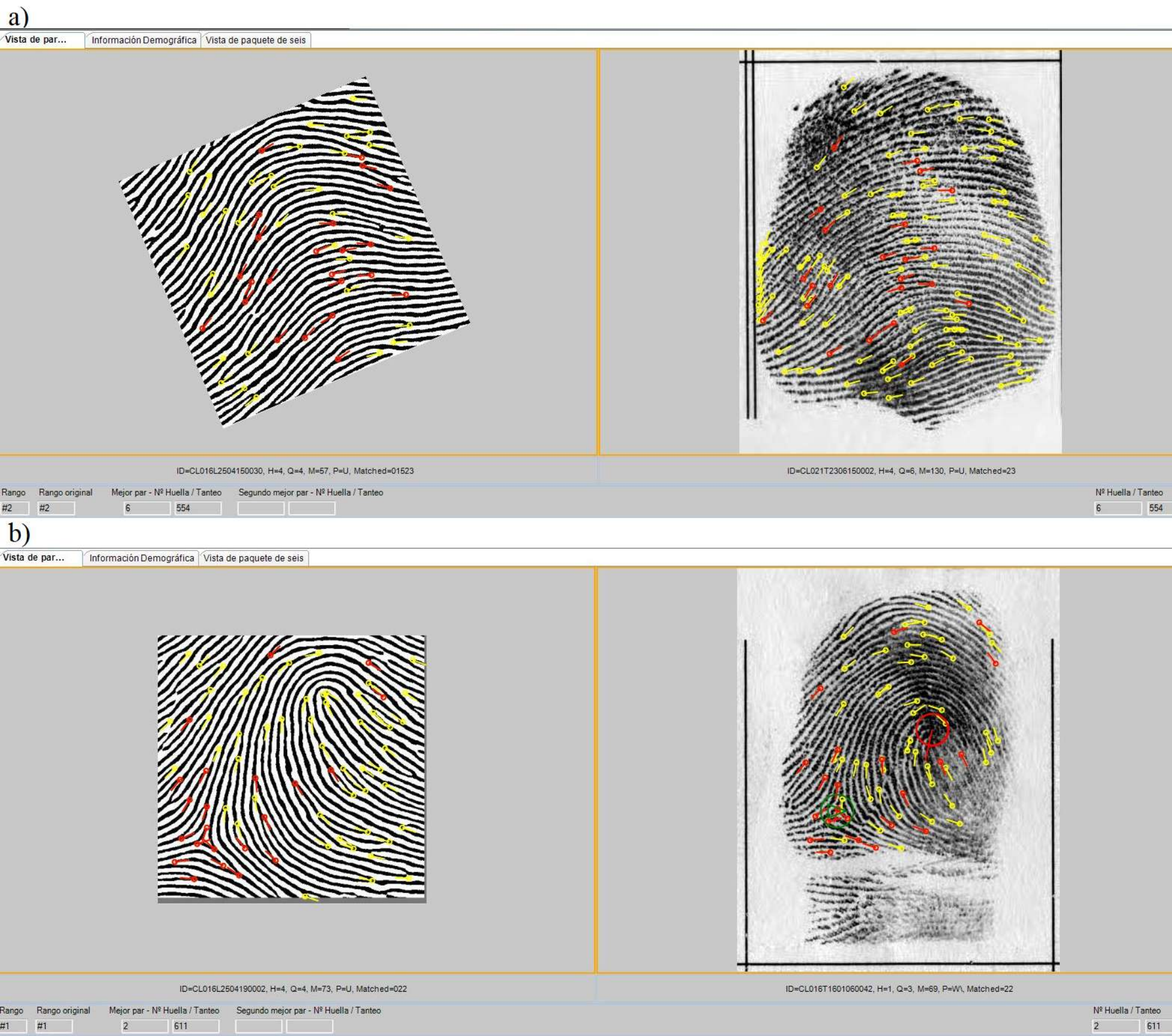}
    \caption{Synthetic fingerprints tested on the PDI database using the AFIS computational program. a) corresponds to the test of a synthetic simple arch, and b) corresponds to the test of a synthetic simple external loop. In the fingerprint designated for search, the position and direction of minutiae are marked (with a circle attached to a segment), shown on the left panels of a) and b) as yellow and red circles with segments. The right panels in a) and b) correspond to real fingerprints that AFIS proposed as possible candidates for identification. The red segments on the left panels indicate positive matches in position and direction with the reference on the right panel, while yellow segments indicate mismatches.}
    \label{Afis}
\end{figure}

To illustrate this process, Fig. \ref{Afis} shows some of the tests performed with synthetic fingerprints using AFIS. In the proposed image for analysis, the positions and directions of minutiae are marked (without discriminating the type of minutiae, which is the responsibility of forensic experts), and AFIS compares the given minutiae against its database, producing a list of candidates ordered by likelihood. Although this method requires expert validation to determine the actual match, due to its lack of minutiae classification and dependence on database size, it provides a fast and effective shortlist of possible candidates. These results demonstrate how a computational system like AFIS can process synthetic fingerprints as if they were real, identifying them as plausible fingerprint-like patterns.

\section{Future work and possible applications}

As mentioned earlier, the task of identifying individuals through fingerprint recognition is the result of more than 200 years of study. One of the key milestones in establishing certainty in identification is the combinatorial result attributed to Balthazard, supported by Locard in \cite{locard1923manuel} and experimentally estimated by Galton in \cite{galton}, as presented in the introduction. This milestone states that the probability of similarity based on the coincidence of $n$ minutiae can be approximated as $P(n) = \dfrac{1}{4^{n}}$, based on the assumption that minutiae can be modeled as uniformly distributed stochastic variables.

\begin{figure}[ht]
    \centering
    \includegraphics[width=0.7\textwidth]{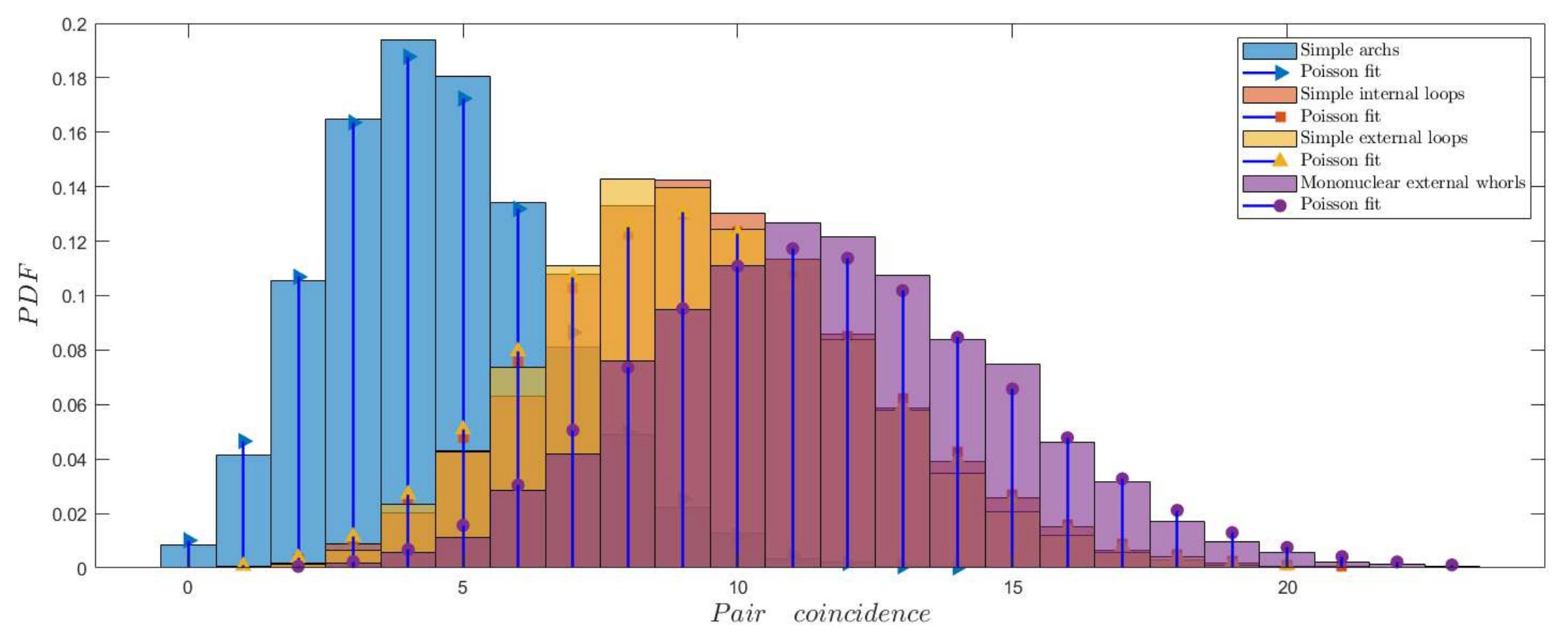}
    \caption{Histogram of the similarity of pairs on synthetic fingerprints done over 404 different fingerprints distributed evenly between simple arches, normal internal loops, normal external loops and mononuclear external whorls as described on section \ref{sec:Validez}. To each histogram of similarity of pairs, a Poisson distribution was adjusted to the data.}
    \label{Poisson}
\end{figure}

Balthazard's hypothesis states that for $n$ pre-established positions, only one of two possible topological accidents---bifurcation or termination---can occur, and these are independent from each other. Every minutiae shown in Fig. \ref{minutas} can be decomposed into these two microstructures. This condition doubles the possible outcomes of the stochastic variables, since both termination and bifurcation have only one axis of symmetry, which is parallel to the directional map. As Balthazard's hypothesis assumes that the $n$ minutiae are independent, the probability of finding $n$ coinciding minutiae becomes the product of the individual probabilities of coincidence. This result is one of the foundational arguments that have led police forces around the world to require more than 10 minutiae for identification. In the specific case of the Chilean identification system, the minimum threshold is 12 minutiae, based on Locard's reasoning \cite{locard1923manuel}. Using the data presented in Section \ref{sec:Validez}, and by categorizing the different structures present in each independent rectangle of the image, it was possible to build a histogram, shown in Fig.~\ref{Poisson}, of similarity based on the coincidence of minutiae positions.

From Fig.~\ref{Poisson}, it is possible to infer that the pairwise coincidence of minutiae, based on the discretization of the image into 512 independent rectangles, is in good agreement with a Poisson distribution. This distribution arises as a limiting case of the binomial distribution when the number of trials is large and the probability of success is small. It can be obtained as the sum of independent binomial variables, which opens the possibility of studying Balthazard's hypothesis from a probabilistic perspective in a PDE model.

The results show a monotonous increase in the $\lambda$ parameter of the Poisson distribution for minutiae pairs, depending on the number of centers and deltas. This trend can be explained by the high concentration of minutiae density around centers and deltas, as illustrated in Fig.~\ref{zonas}. 

The model presented here will be used in future work to explore the relationship between minutiae coincidence in synthetic fingerprints and extrapolate these findings to real fingerprints. This approach provides a more robust mathematical foundation for fingerprint analysis, helping to better understand the principles behind Balthazard's hypothesis and to compare it with more modern models.

\medskip 
It is also worth mentioning that in GenCHSin we have developed a feature to make simulations resemble real fingerprints more closely. This involves subdividing the pattern into independent rectangles---similar to the subdivisions shown in Fig.~\ref{zonas}, but containing no more than one synthetic ridge each---so they can be manipulated as independent domains capable of hosting pores along the ridge. The positions of the pores $(x_{p},y_{p})$ are selected stochastically, following a uniform distribution on the synthetic ridge inside each independent rectangle, and the radius of each pore $r_{p}$, which is measured in pixels, is selected following a Gaussian distribution centered at $3$ with a standard deviation of $0.7$. The discretized domain in the interior of each synthetic pore, which corresponds to the family of pixels $(x,y)$ that satisfy $(x - x_{p})^{2} + (y - y_{p})^{2} \leq r_{p}^{2}$, is set to the maximum intensity value in the standard grayscale range, which corresponds to 255.

\begin{figure}[ht]
    \centering
    \includegraphics[width=0.20\textwidth]{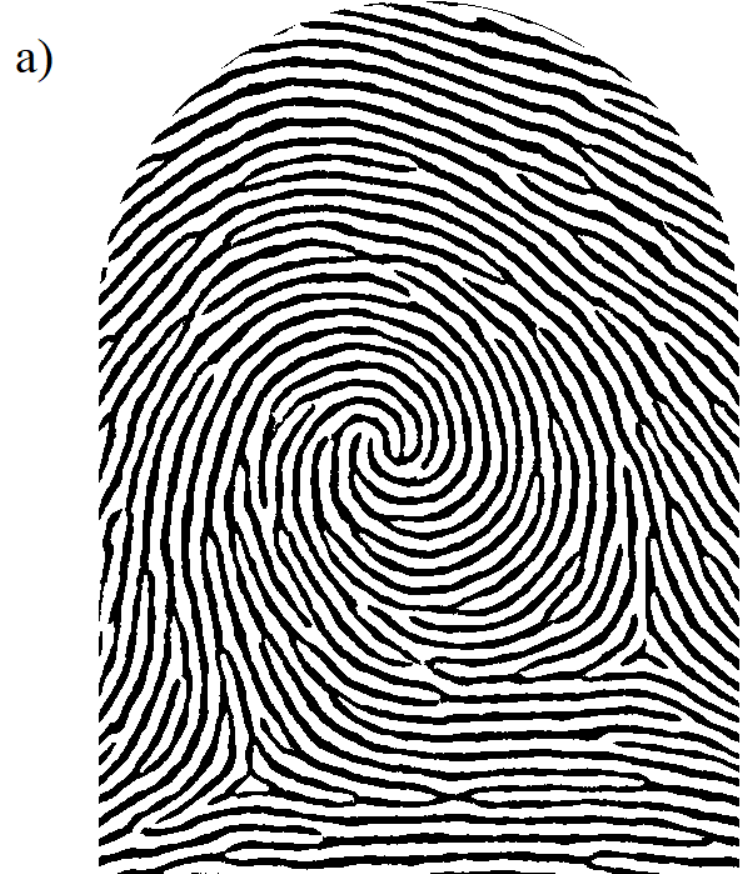}
    \includegraphics[width=0.20\textwidth]{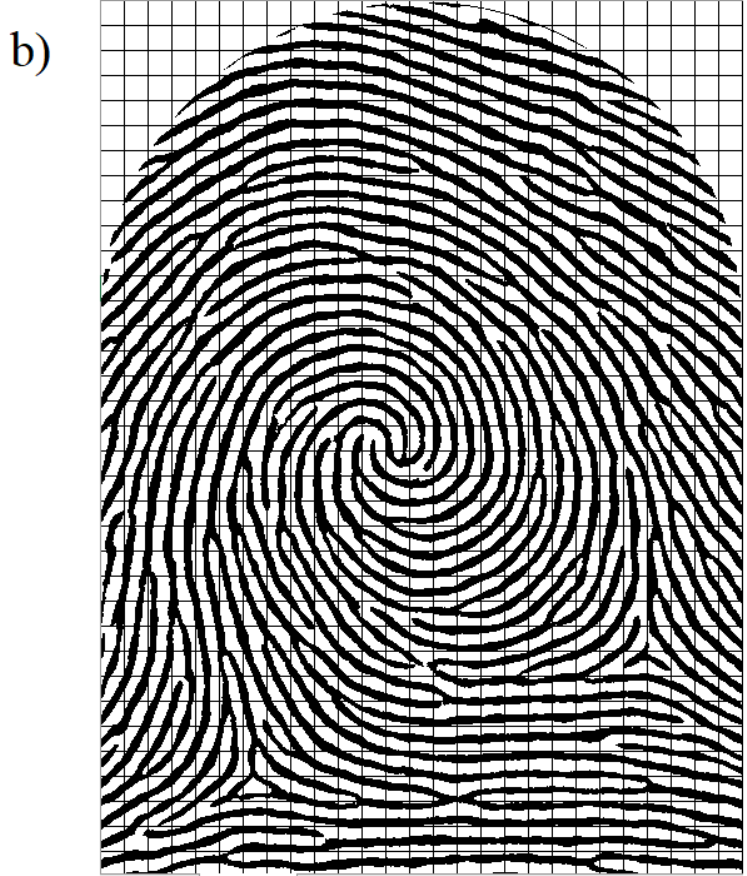}
    \includegraphics[width=0.20\textwidth]{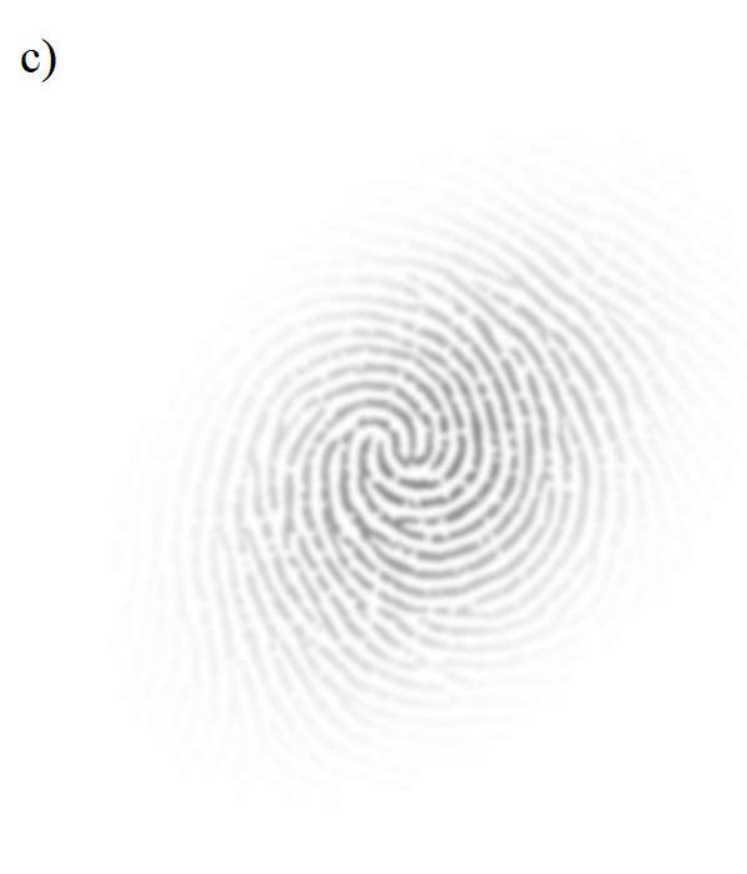}
    \caption{a) Synthetic fingerprint generated with GenCHSin before applying the porous fixation process described. b) Synthetic fingerprint showing the grid used to define the porous sections. c) Final result given by GenCHSin after applying the porous fixation, Gaussian filtering, and blurring process.}
    \label{Poros}
\end{figure}

Once all pores are placed, a Gaussian filter is applied, centered on the nuclear region of the synthetic fingerprint, so that the intensity of the pattern decays exponentially with distance from the center. Finally, the resulting pattern is blurred to simulate the more realistic appearance of a fingerprint that could be recovered from a scene. An example of this process is shown in Fig.~\ref{Poros}.

The structure and distribution of pores must be analyzed to assess the degree of similarity between the final output of GenCHSin and real fingerprints.

\medskip
To conclude this section, we remark that the model's characteristics make it suitable for addressing open questions regarding the reliability of fingerprint analysis, the independence of minutiae appearance, and the minimum number of minutiae coincidences required in different countries for human identification via fingerprints. Moreover, its good agreement with real fingerprint characteristics, the wide range of parameters, and the stochastic nature of the initial conditions provide the possibility of applying this model in industrial contexts, particularly in biosynthetic security as a complement to biometric security in both physical and digital forms.

\section{Conclusions}

We have successfully built a model that can fully generate different kinds of fingerprint pattern formed on humans, developing the open source software named GenCHSin that is based on the specific Chilean classification key, and can exhibit the wide variety of minutiae and ridge organization which can be seen on real fingerprints. The model presented here has the characteristic of having multiple stationary states which only depend on the initial conditions taken in every simulation, which gives the possibility of generating a large amount of different synthetic fingerprints which have common parameters.

The model presented is in good agreement with the different characteristics exhibited by real fingerprint, such as ridge direction, border conditions, number of minutiae per fingerprint, ridge density and characterization of zones on the domain. The resulting images can be analyzed by different identification software, such as AFIS, which is used on human identification via fingerprint recognition, determining its different characteristics.

This model has the potential to open a wide variety of aspects in fingerprint applications, such as the construction of complementary biosynthetic methods of identification, virtual identities or the construction of training databases for artificial intelligence programs that do not violate the privacy of people, and a wide variety of aspects in fingerprint analysis, such as the construction of similarity models that can ensure a minimal analysis for human identification.

\section*{Acknowledgments}
C.R. was supported by ANID FONDECYT 1231593.
A.O. was partially funded by ANID-Fondecyt 1240200, 1231404, 
CMM FB210005 Basal-ANID, FONDAP/1523A0002, FONDEF IT23I0095, 
ECOS240038 and DO210001. 


\bibliography{FingerprintsReferences}
\end{document}